\documentclass{article}
\usepackage[utf8]{inputenc}
\usepackage{url}
\usepackage{algorithm}
\usepackage{algpseudocodex}
\usepackage{graphicx,algorithm,algcompatible,amsmath, relsize}
\usepackage{hyperref}
\usepackage{subfigure}
\usepackage{mathrsfs}
\usepackage[utf8]{inputenc}
\usepackage[T1]{fontenc}
\usepackage{amsthm,amssymb,inputenc,titling}
\usepackage{hyperref}
\usepackage[left=3.5cm, right=3.5cm, top=1.5cm, bottom=2cm]{geometry}
\title{\textbf{Multi-Document Financial Question Answering using LLMs}}
\author{Shalin Shah\\ \small Anvai AI\\ \small shalin@anvai.ai 
        \and
        Srikanth Ryali\\ \small Anvai AI\\ \small srikanth@anvai.ai
        \and
        Ramasubbu Venkatesh\\ \small Anvai AI\\ \small venky@anvai.ai
        }\date{}
\begin{document}
\maketitle
\begin{abstract}
We propose two new methods for multi-document financial question answering. First, a method that uses semantic tagging, and then, queries the index to get the context (RAG\_SEM). And second, a Knowledge Graph (KG\_RAG) based method that uses semantic tagging, and, retrieves knowledge graph triples from a graph database, as context. KG\_RAG uses knowledge graphs constructed using a small model that is fine-tuned using knowledge distillation using a large teacher model. The data consists of 18 10K reports of Apple, Microsoft, Alphabet, NVIDIA, Amazon and Tesla for the years 2021, 2022 and 2023. The list of questions in the data consists of 111 complex questions including many esoteric questions that are difficult to answer and the answers are not completely obvious. As evaluation metrics, we use overall scores as well as segmented scores for measurement including the faithfulness, relevance, correctness, similarity, an LLM based overall score and the rouge scores as well as a similarity of embeddings. We find that both methods outperform plain RAG significantly. KG\_RAG outperforms RAG\_SEM in four out of nine metrics.
\end{abstract}
\section{Introduction}
LLMs are now increasingly used for various linguistics tasks like information extraction \cite{peng2024metaie} \cite{xu2023large} \cite{peng2023embedding}, language understanding \cite{dada2024clue} \cite{karanikolas2023large} , named entity recognition \cite{wang2023gpt} \cite{monajatipoor2024llms} \cite{hu2024improving}, text summarization (including query based summarization) \cite{shah_topic} \cite{edge2024local} \cite{jin2024comprehensive}, translation \cite{gao2024llms} \cite{enis2024llm}, speech synthesis \cite{li2024whisma}, question answering \cite{arefeen2024leancontext} \cite{zhuang2024toolqa} \cite{prabhu2024dexter} \cite{bhat2023investigating}, code generation/understanding \cite{nam2024using}, and other computational linguistics tasks. In this work, we focus our attention on financial question answering. Specifically multi-document financial question answering. We use 10k reports from six companies for three years. This is different from asking questions about one document specific to a year, domain and organization, which is a significantly easier problem. Our method, which uses semantic tagging and knowledge graphs can scale to a very large corpus of documents including multiple publication dates, multiple domains, and multiple industries.\\\\
Financial question answering using LLMs has been recently studied \cite{srivastava2024evaluating} \cite{lee2024survey} and many attempts have been successful and widely used. There are a few benchmark datasets for financial linguistics \cite{xie2024finben} \cite{xie2024pixiu} \cite{zhuang2024toolqa} \cite{prabhu2024dexter} but most offer single document question answering. In this paper, we propose two new methods, RAG\_SEM and KG\_RAG for multi-document question answering. We perform experiments using 18 10k reports, spanning three years and 111 questions.\\\\
Multi-document question answering has been studied recently (including methods based on knowledge graphs) \cite{wang2024knowledge} but the exploration of this subject has been limited. Knowledge graphs are very suitable for multi-document question answering because of the fine-grained semantics that they model. Works such as \cite{edge2024local} could be explored for this purpose.\\\\
Specialized models fine-tuned on large financial text corpuses have been shared for the general public to consume \cite{wu2023bloomberggpt} \cite{liu2023fingpt} \cite{lee2024survey}. However, their use has been limited due to the risk of hallucinations, that increases for models trained using transfer learning \cite{gekhman2024does}. We do not fine-tune models that generate answers, but we do fine-tune a KG generator, for speed and cost saving.\\\\
In this work, we introduce two new methods for multi-document financial question answering. Multi-document question answering is difficult because of the fact that the context (RAG) is generated from all documents rather than a document specific to a domain, industry or organization. This makes it challenging as the synthesizer LLM may start giving wrong answers when asked a question about a specific entity. For example, if the question is "How are companies' revenues in the finance industry?" might retrieve information about closely related topics, like economics, business, investing, insurance or lending. This may impact the overall effectiveness of the synthesized response.
\begin{figure}[ht]
\label{comparison_1}
\centering
\includegraphics[scale=0.6]{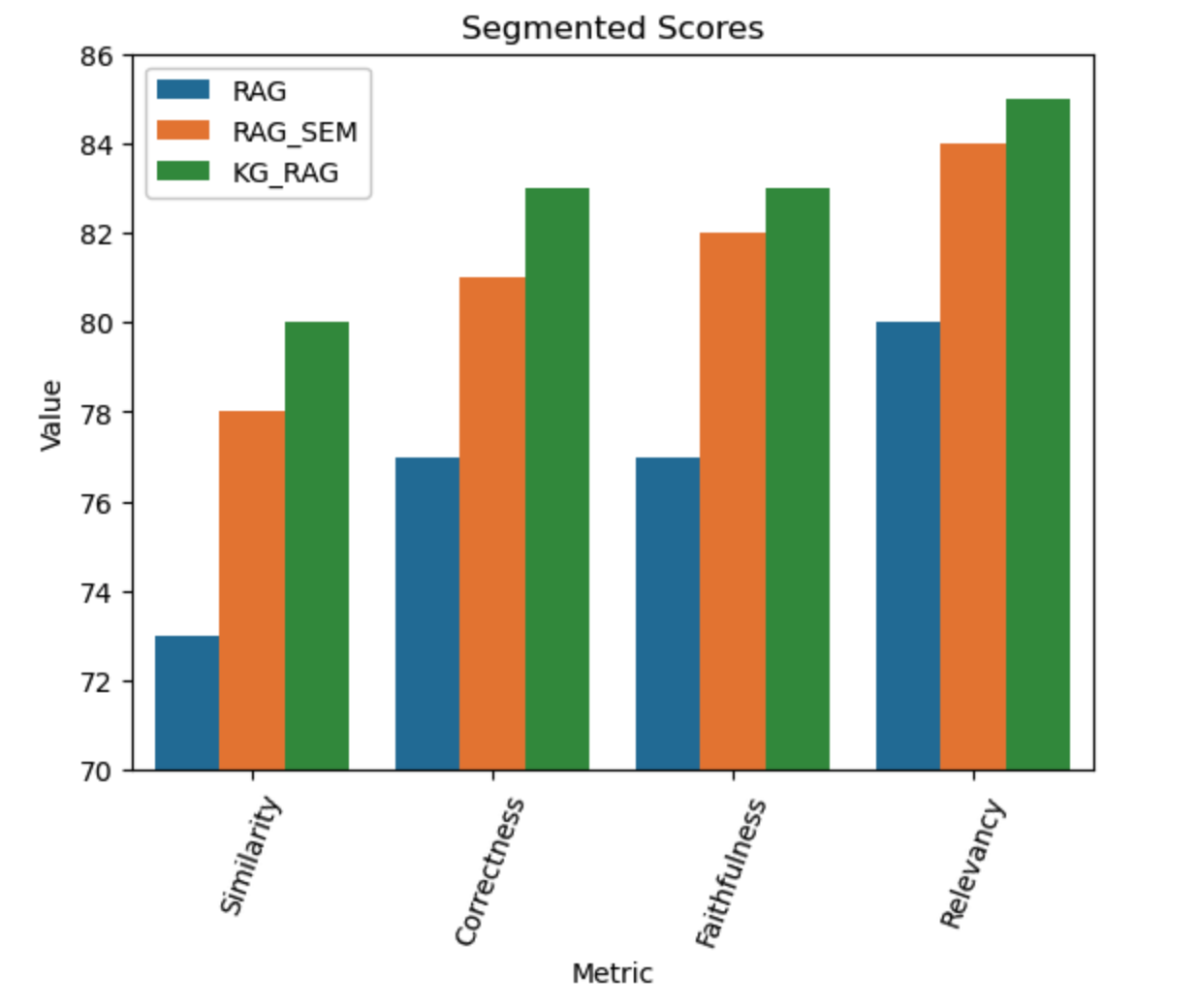}
\caption{Segmented Comparison of the Three Methods of Financial Question Answering}
\end{figure}

\begin{figure}[ht]
\label{comparison_2}
\centering
\includegraphics[scale=0.6]{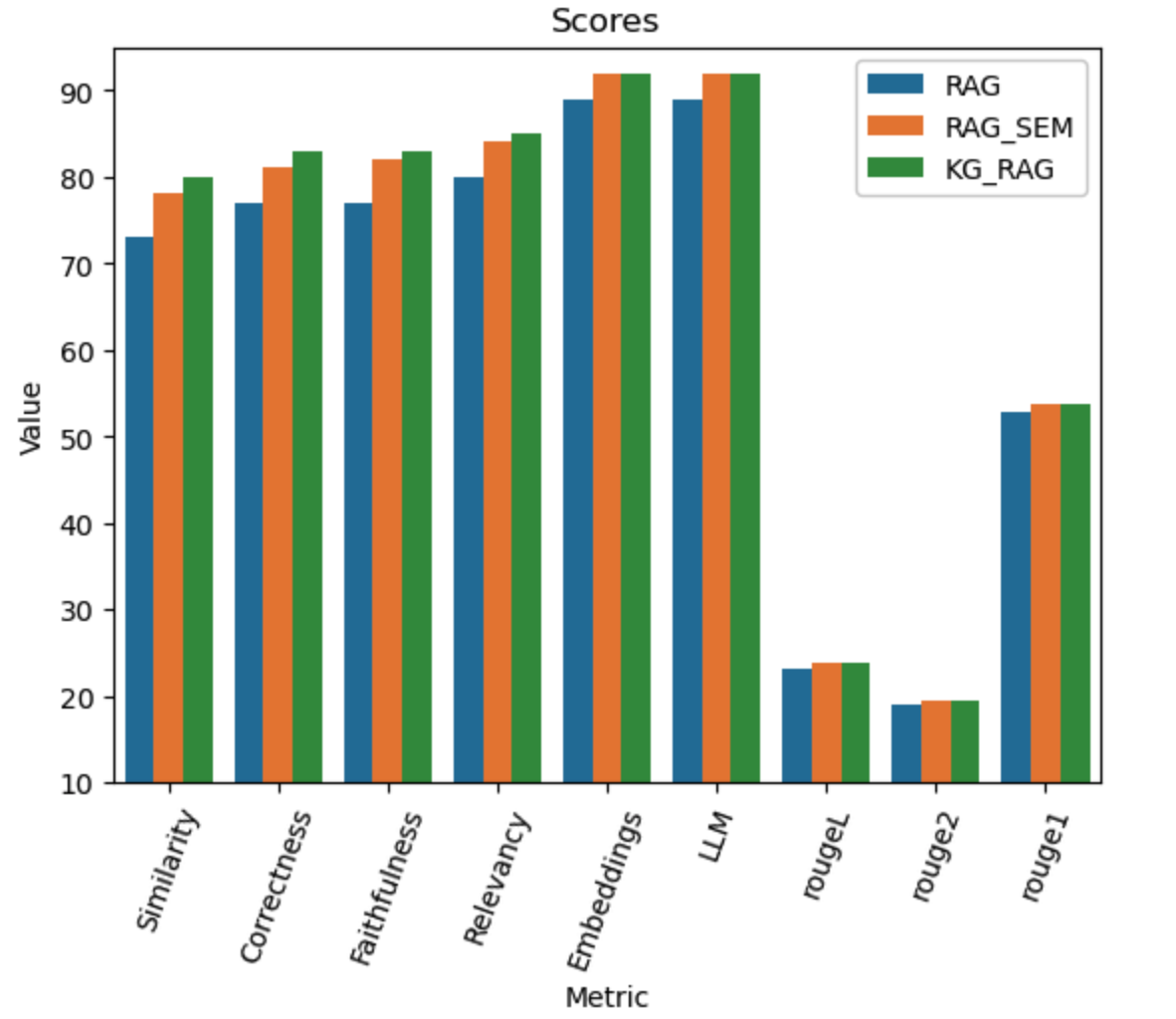}
\caption{Comparison of the Three Methods of Financial Question Answering}
\end{figure}

\begin{table}[t]
\caption{Experiment Results (18 10K Reports and 111 Questions)}
\label{results1}
\begin{center}
\begin{tabular}{p{2cm}p{2cm}p{2cm}p{2cm}}
\hline
\bf Metric &\bf RAG &\bf RAG\_SEM &\bf KG\_RAG\\
\hline
Relevance &80\% &84\% &\bf 85\%\\
\hline
Correctness &77\% &81\% &\bf 83\%\\
\hline
Faithfulness &77\% &82\% &\bf 83\%\\
\hline
Similarity &73\% &78\% &\bf 80\%\\
\hline
LLM Score &89\% &\bf 92\% &\bf 92\%\\
\hline
Embeddings &89\% &\bf 92\% &\bf 92\%\\
\hline
rouge1 &52.9\% &\bf 53.8\% &\bf 53.8\%\\
\hline
rouge2 &19\% &\bf 19.5\% &\bf 19.5\%\\
\hline
rougeL &23.2\% &\bf 23.8\% &\bf 23.8\%\\
\hline
\end{tabular}
\end{center}
\end{table}
\section{Proposed Methods}
The knowledge of an LLM depends on a very large data set that is used to train it. Furthermore, the training is done using techniques like masking and next word prediction. The knowledge of an LLM is a highly compressed form of Terabytes of data. LLMs can reason and generate language effectively, but needle in a haystack problems are difficult for an LLM to solve. An LLM needs context on the basis of which the answer should be generated, in addition to the world knowledge that the LLM learns. There are methods like retrieval augmented generation (RAG) \cite{lewis2020retrieval} as well as few shot learning \cite{brown2020language} which have been successfully used in several domains like medicine, finance, law and so on.\\\\
However, in the case in which there are multiple sources of information, and the answer should use only a few of the sources, semantic tagging and agentic pipelines are highly useful. We use semantic tagging, which tags a piece of text with tags like named entities, dates, industries, domains and sectors, organizations, partnerships, dividends etc.\\\\
In addition to semantic tagging and retrieval augmented generation, we use knowledge graph RAG which adds knowledge graph triples (or the related synthesized text) to the context. Knowledge graph RAG has been somewhat explored before in works like \cite{sanmartin2024kg} \cite{matsumoto2024kragen} \cite{sarmah2024hybridrag}.\\\\
We propose two new methods, RAG\_SEM and KG\_RAG which are methods that enhance a RAG system using semantic tagging and knowledge graphs. We describe both methods below, and also in figures 3, 4 and 5. First we give a short introduction to retrieval augmented generation and semantic tagging (sections 3.1 and 3.2). Sections 3.3 and 3.4 describe our methods.
\subsection{RAG\_SEM}
Our method RAG\_SEM is show in figure 4, and algorithm 2. The indexing phase of this method tags each of the 18 10k reports with semantic tags like industries, organization names, dates, partnerships, products and so on. We use an LLM to generate the semantic tags for documents and questions, using a custom prompt.\\\\
Then, as in a traditional RAG system, we semantically chunk all 18 10k reports and store the chunks in a vector database, along with the tags.\\\\
When a question is asked, we first tag the question using the same prompt that was used to tag the documents. We pass the context and the tags along with a custom synthesis prompt to the LLM and generate the final response.\\\\
This has several benefits in our case where there are several organizations and several years of documents. The tags help, in some way, to choose the right parts of the index.
\subsection{KG\_RAG}
In addition to semantic tagging and the retrieval of chunks, this method generates knowledge graph triples from the documents. Figure 5 and algorithm 3 succinctly describe the method. KG triples are highly fine-grained, and can be used for effective retrieval of needle in a haystack facts. Furthermore, KG triples are short pieces of facts, and can be used in a large number without burdening the context length limits of an LLM. But, even if long context length LLMs like Gemini \cite{team2023gemini} are available, passing in too much text to an LLM in a RAG pipeline might degrade the quality of the response.\\\\
Knowledge graphs give structure to unstructured text and help in question answering that requires multi-hopping. In addition to the retrieved nodes, neighborhoods are also extracted that increases the semantics of the retrieved context.\\\\
In KG\_RAG, after semantic tagging, we retrieve chunks and KG triples and pass both to an LLM through a custom prompt. Results in table 1 and figures 1 and 2 show that this method outperforms RAG in all nine metrics and outperforms RAG\_SEM in four out of nine metrics.\\\\
And advantage of KG\_RAG is that it uses multi-hop question answering. This retrieves nodes that match the question as well as nodes in the immediate neighborhood of the initially selected nodes. Neighborhood propagation of the textual information might also help, as described in \cite{edge2024local}.
\subsection{Knowledge Graphs}
A knowledge graph (KG) is a graph with nodes and edges, and it represents a set of triples:\\\\
$(Subject \rightarrow Predicate \rightarrow Object)$\\\\
Examples:
\begin{itemize}
    \item $(GoogleCloud \rightarrow OperatingIncome \rightarrow \$1.7B)$
    \item $(Alphabet \rightarrow R\&DInvestments \rightarrow \$45B)$
    \item $(Microsoft \rightarrow Acquired \rightarrow ZenimaxMedia)$
    \item $(Microsoft \rightarrow WindowsOEMRevenue \rightarrow 25\%Decrease)$
\end{itemize}
\subsection{Knowledge Graph Extraction using LLMs}
Construction of knowledge graphs using an LLM has not been adequately explored \cite{zhu2024llms} \cite{meyer2023llm} \cite{kommineni2024human}. We develop a large custom prompt to generate knowledge graph triples, which includes detailed instructions on what we are looking for e.g. "Pay special attention to trigger words like increased, decreased, and reduced".\\\\
Furthermore, we fine-tune a small language model using knowledge distillation from a large language model to construct knowledge graphs. The small model responds several orders of magnitude faster and it also requires less compute power.
\subsection{A Sample of the Financial Questions we ask the LLM}
We list 4 questions below. Most questions are difficult esoteric questions that regular approaches such as RAG cannot answer:
\begin{enumerate}
\item What insights can be drawn from the 10-K disclosures of Apple, Microsoft, and Alphabet about their leadership development programs, succession planning, and management training initiatives, and how do these initiatives align with their broader strategic objectives?
\item How does Apple’s commitment to achieving carbon neutrality across its supply chain and products by 2030, as discussed in its 10-K, affect its cost structure, supplier relationships, and long-term profitability, and what are the potential risks and rewards associated with this aggressive ESG strategy?
\item How does Alphabet’s allocation of resources and capital to its “Other Bets” segment (like Waymo, Verily, and X) reflect its strategic vision for diversification beyond its core advertising business, and what are the potential risks and rewards associated with this portfolio approach?
\item What are the most significant risks Amazon highlights, and how might these impact its competitive positioning in key markets like cloud computing, e-commerce, and digital advertising?
\end{enumerate}
\begin{figure}[ht]
\label{rag}
\centering
\includegraphics[scale=0.5]{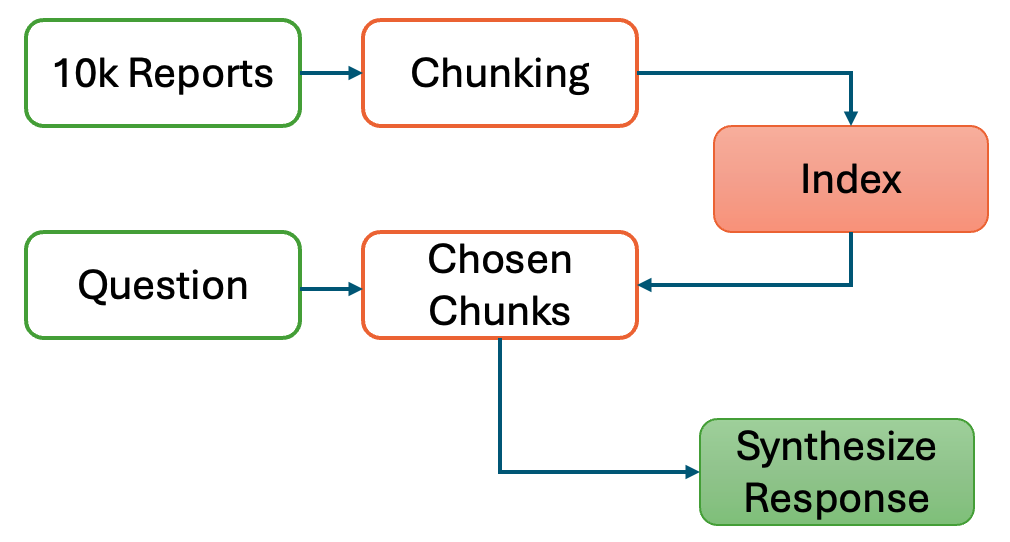}
\caption{\textbf{RAG}: Plain Retrieval Augmented Generation}
\end{figure}
\begin{figure}[ht]
\label{rag_sem}
\centering
\includegraphics[scale=0.5]{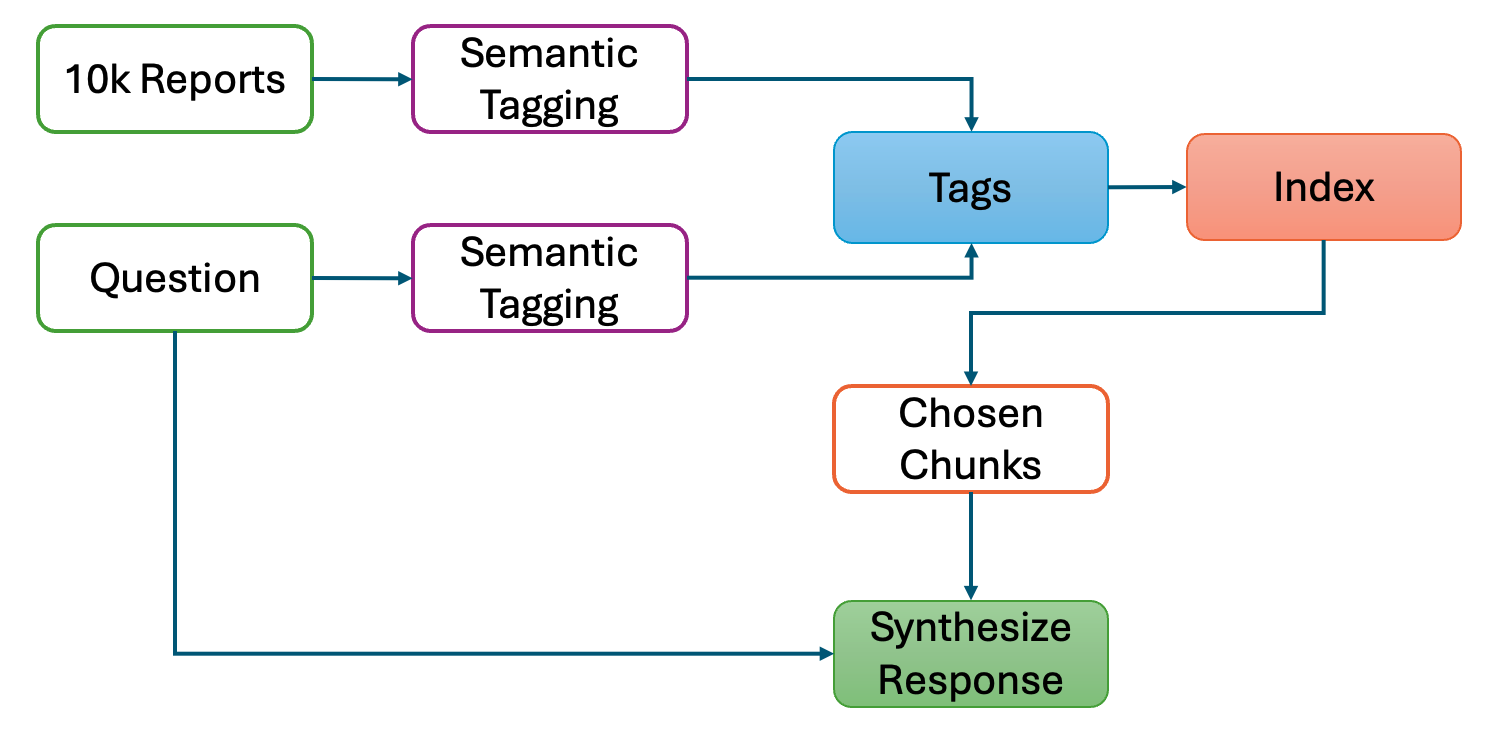}
\caption{\textbf{RAG\_SEM}: RAG with Semantic Tagging}
\end{figure}
\begin{figure}[ht]
\label{kg_rag}
\centering
\includegraphics[scale=0.5]{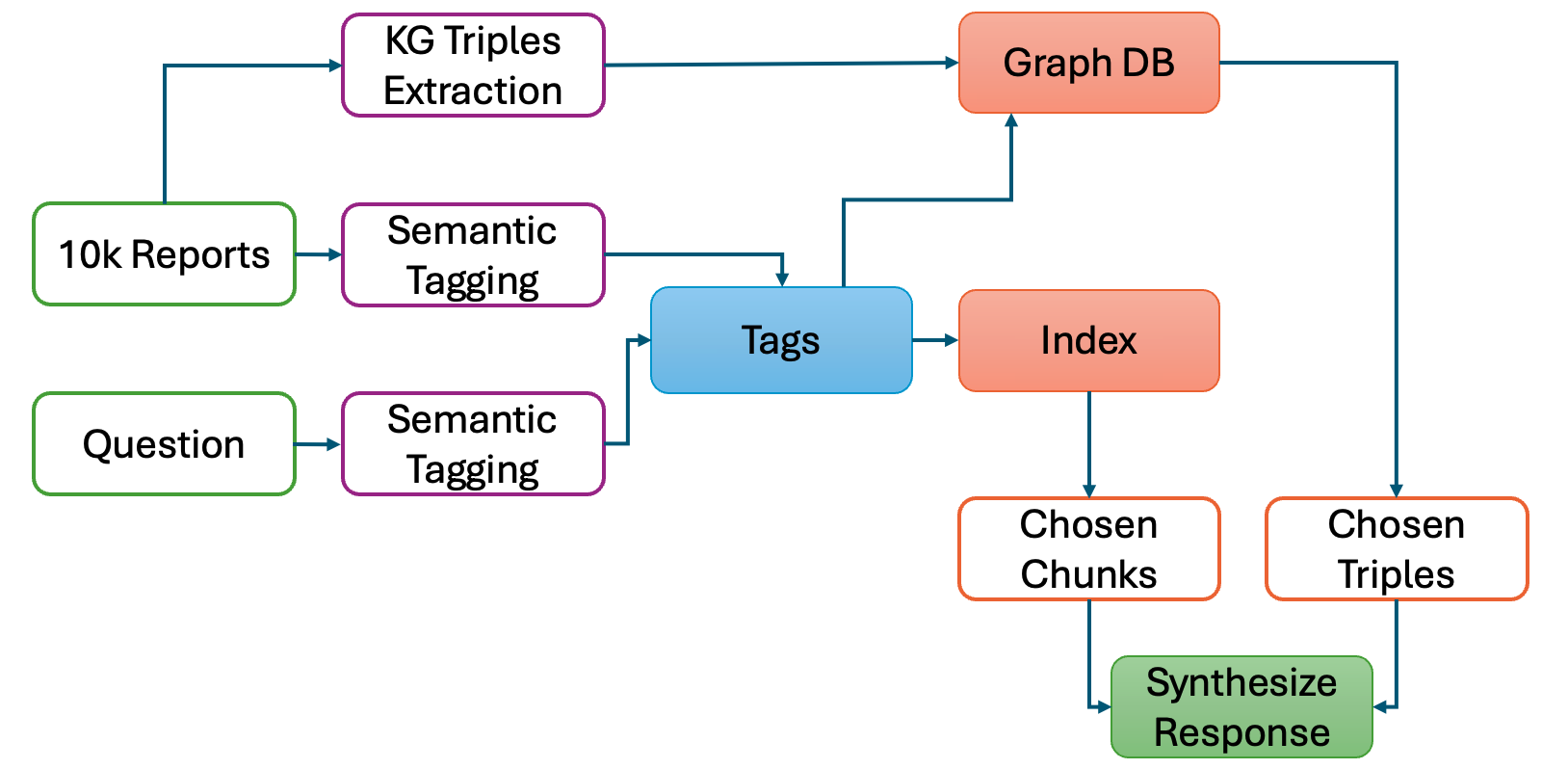}
\caption{\textbf{KG\_RAG}: Knowledge Graph RAG with Semantic Tagging}
\end{figure}
\begin{algorithm}
\caption{\textbf{RAG}}
\textbf{Input}: A Question, \textbf{Output}: The Answer
\begin{algorithmic}[1]
\State Send the question to a vector database
\State Retrieve chunks
\State Generate prompt
\State Send the question, the prompt and the retrieved chunks to an LLM
\State Synthesize the response
\end{algorithmic}
\end{algorithm}
\begin{algorithm}
\caption{\textbf{RAG\_SEM}}
\textbf{Input}: A Question, \textbf{Output}: The Answer
\begin{algorithmic}[1]
\State Semantic tagging of the question
\State Retrieve chunks
\State Generate prompt
\State Send the question, the prompt and the retrieved chunks to an LLM
\State Synthesize the response
\end{algorithmic}
\end{algorithm}
\begin{algorithm}
\caption{\textbf{KG\_RAG}}
\textbf{Input}: A Question, \textbf{Output}: The Answer
\begin{algorithmic}[1]
\State Semantic tagging of the question
\State Retrieve chunks
\State Retrieve knowledge graph triples
\State Generate prompt
\State Send the question, the prompt, the retrieved chunks and knowledge graph triples to an LLM
\State Synthesize the response
\end{algorithmic}
\end{algorithm}
\section{Experiments and Results}
Table 1 and figures 1 and 2 show the results. \textbf{Both of our models} outperform vanilla RAG on all nine metrics, and the \textbf{knowledge graph RAG} outperforms semantic RAG on four out of nine metrics.\\\\
We evaluate RAG, RAG\_SEM and KG\_RAG and show the results in table 1, as well as figures 1 and 2.\\\\
We use nine metrics for our evaluation:
\begin{itemize}
    \item \textbf{rouge scores} \cite{lin2004rouge}: These scores are n-gram level metrics, and rouge1 uses unigrams, rouge2 uses bigrams and rougeL uses a longest common subsequence (rouge scores use stemming, so exact matches are not required).
    \item \textbf{Faithfulness} \cite{es2023ragas}: how factually accurate the answer is.
    \item \textbf{Correctness} \cite{es2023ragas}: a combined score of the semantic relatedness and factual similarity.
    \item \textbf{Relevancy} \cite{es2023ragas}: a score of how relevant an answer is to the question.
    \item \textbf{Similarity} \cite{es2023ragas}: a score of how semantic similarity
    \item \textbf{LLM Scoring}: ask the LLM to evaluate the test answers and generate an overall score
    \item \textbf{Embeddings}: compute a cosine similarity
\end{itemize}
For metrics 2, 3, 4 and 5, which are described in \cite{es2023ragas}, we pass in the entire text, the question, the reference answer, the generated answer from the test model and the definitions of the metrics, to an LLM for evaluation. This gives the LLM a comprehensive context on the basis of which to evaluate the control (RAG) and the test (RAG\_SEM and KG\_RAG) models.
\section{Key Contributions}
\begin{itemize}
    \item We study the problem of multi-document financial question answering which is different and significantly more difficult the single document domain-specific question answering
    \item We use 18 10k reports (Apple, Microsoft, NVIDIA, Amazon, Tesla and Alphabet for years 2021, 2022 and 2023)
    \item Our method is horizontally scalable, with possibly no performance impact for adding more documents, possibly in different domains, industries and sectors
    \item Both of our proposed methods extract semantic tags from the documents as well as the posed question, such as named entities, dates, industries, locations, partners, and dividends.
    \item Our method RAG\_SEM uses semantic tagging with retrieval augmented generation (RAG) to generate answers
    \item Our method KG\_RAG uses semantic tagging along with RAG and knowledge graphs to synthesize answers
    \item We generate knowledge graphs using a small LLM fine-tuned using knowledge distillation from a large LLM.
    \item We extensively evaluate both proposed methods against plain retrieval augmented generation (RAG) on nine metrics (rouge1/rouge2/rougeL scores, faithfulness, correctness, relevance, similarity, LLM scoring and embedding similarity)
    \item RAG\_SEM outperforms RAG on all nine metrics. KG\_RAG outperforms RAG on all nine metrics. KG\_RAG outperforms RAG\_SEM on four metrics.
\end{itemize}
\section{Conclusion}
We proposed two new methods RAG\_SEM and KG\_RAG that outperform vanilla RAG in all nine metrics. Furthermore, KG\_RAG outperforms RAG\_SEM on four out of nine metrics. These four metrics are segmented scores, such as faithfulness, correctness, relevance and similarity \cite{es2023ragas}. Future work could be a full agentic routing instead of retrieving the context using semantic tagging. Comparisons could include methods like GraphRAG \cite{edge2024local} for the metrics as well as the latency.\\\\
Here are the main conclusions:
\begin{itemize}
    \item \textbf{Superiority of the Methods}: The two methods demonstrate superior performance over the control model across all nine evaluated metrics in the context of financial question answering. This consistent outperformance suggests that our proposed methods based on LLMs have a greater capacity to understand and generate accurate responses to financial queries, highlighting their effectiveness and robustness in this domain.
    \item \textbf{Implications for Financial Applications}: The findings indicate that advanced methods like KG\_RAG can be more effective than traditional or simpler models for financial question answering tasks. This suggests significant potential for the adoption of such models in practical applications, such as financial advisory services, automated customer support, and decision support systems in financial institutions. The improved performance across diverse metrics also emphasizes the versatility of LLMs in handling various aspects of financial information processing, from accuracy and relevance to contextual understanding.
    \item \textbf{Future Research Directions}: Given the promising results, future research could explore further optimization of the methods, including fine-tuning for specific financial sub-domains or incorporating domain-specific knowledge sources like proprietary databases. Additionally, research could focus on understanding the specific factors or model architectures that contribute most to the observed performance gains, as well as evaluating the LLMs on a wider range of financial tasks to generalize these findings.
    \item \textbf{Other applications} of the method will include non-financial applications in domains such as medicine, agriculture, public policy, law, education, retail, supply chain, environmental science, business intelligence, and research.
 \end{itemize}
\nocite{*}
\bibliographystyle{unsrt}
\bibliography{kg}
\section{Appendix: Knowledge Graphs Constructed from Finance Documents}
Figure 6 shows a knowledge graph extracted from a short financial document.\\\\
The knowledge graphs are constructed using a fine-tuned small model, and then visualized using graphviz \cite{ellson2002graphviz}.
\begin{figure}[ht]
\label{kg_1}
\centering
\includegraphics[scale=0.4]{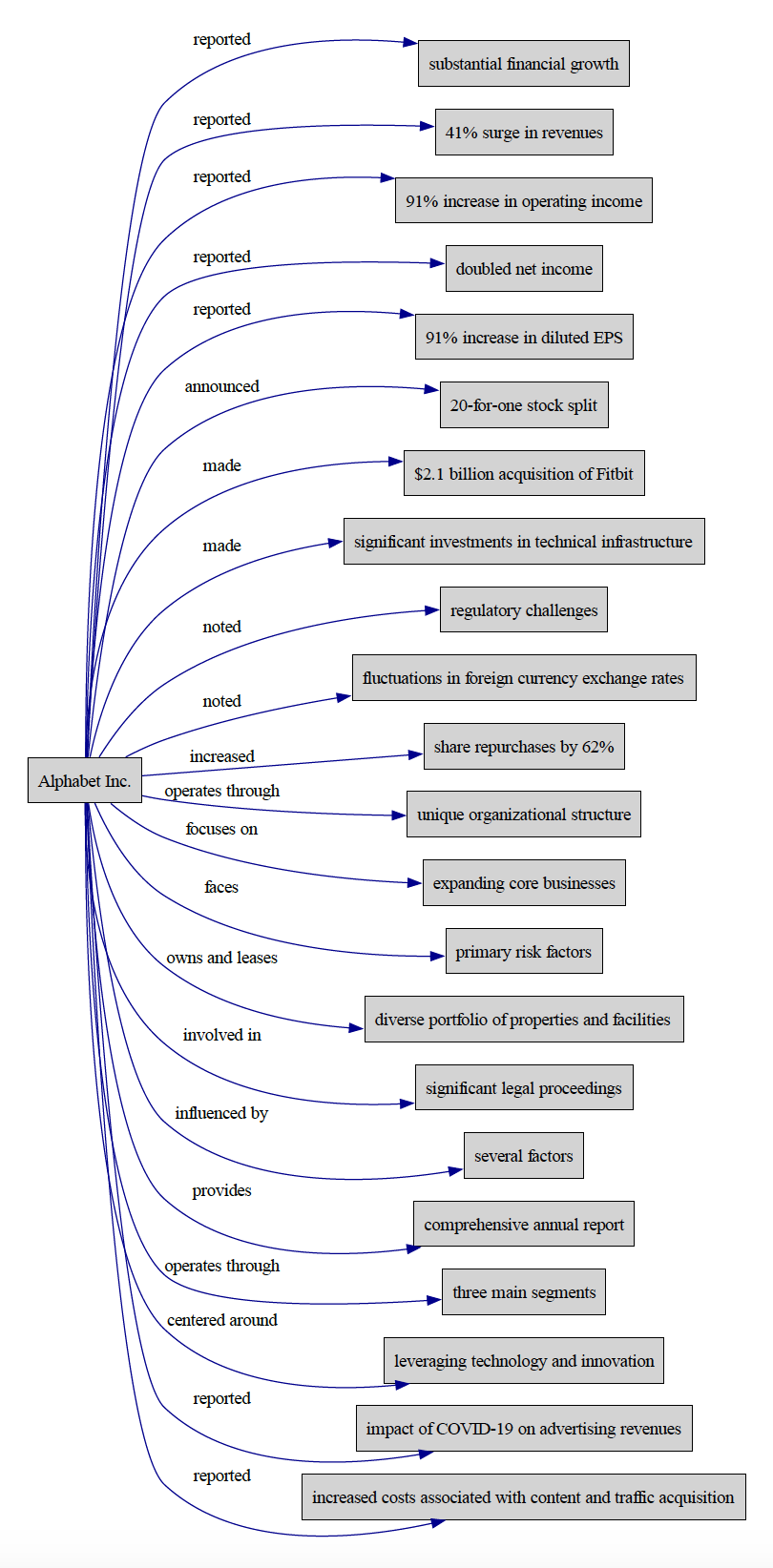}
\caption{A Knowledge Graph Extracted from a Financial Document}
\end{figure}
\end{document}